\documentclass{edp-conf}
\usepackage{graphicx,subfigure}
\begin{document}

\TitreGlobal{Supernova Legacy Survey (SNLS) : real time operations and photometric analysis}

%%-----------------------------
%%      the top matter
%%-----------------------------
\title{Supernova Legacy Survey (SNLS): real time operations and photometric analysis}
\author{Palanque-Delabrouille, N. (on behalf of the SNLS collaboration)}\address{DAPNIA/SPP, CEA Saclay, 91191 Gif-sur-Yvette Cedex, France}
\runningtitle{SNLS: real time operations and photometric analysis}
%
%\setcounter{page}{237}
% Keep this line, even if the page will be settled afterwards..
\index{Palanque-Delabrouille, N.}
% Repeat the authors here, this will help to make the final index

\maketitle
\begin{abstract} 
Type Ia supernovae (SN Ia) have provided the first evidence for an accelerating universe and for the existence of an unknown ``dark energy'' driving this expansion. The 5-year Supernova Legacy Survey (SNLS) will deliver $\sim $700 type Ia supernovae and as many type II supernovae with well-sampled light curves in 4 filters $g'$, $r'$, $i'$ and $z'$. The current status of the project will be presented, along with the real time processing leading to the discovery and spectroscopic observation of the supernovae. We also present an offline selection of the SN candidates which aims at identifying and eliminating potential selection biases.

\end{abstract}
%
%%-----------------------------
%%      your text
%%-----------------------------
\section{Introduction}

Exploding under extremely constrained conditions, type Ia supernovae have been used as standard candles for over a decade to study the evolution of the expansion rate of the Universe. In 1998, two teams (Riess et al. 1998; Perlmutter et al. 1999) independently announced that the expansion of the Universe is accelerating. This discovery ranks as one the major developments in cosmology. Translated in terms of the cosmological parameters, this results in roughly 70\% of the energy density of the Universe composed of a new and unknown form called dark energy which drives the universal expansion. 

This unexpected result has since then been confirmed by independent studies based on the observation of the anisotropies of the cosmic microwave background or on galaxy clusters. The nature of this dark energy component however still remains unknown. The key parameter for its description is the equation-of-state parameter $w$ through the relation between its pressure $p$ and density $\rho$ ($p=w\rho$). A classical cosmological constant $\Lambda$ as introduced by Einstein has $w=-1$ whereas more general quintessence scenarii yield $w>-1$. The determination of the equation of state of the dark energy is therefore crucial in the understanding of its nature, and is one of the main objectives of the Supernova Legacy Survey (SNLS).

\section{Overview of the SNLS experiment}

The SNLS experiment has been designed to confirm the discovery of the accelerating expansion of the universe and improve our knowledge of the dark energy component. Data from the Canada-France-Hawaii Telescope Legacy Survey (CFHTLS) has been cumulated for this purpose since June 2003, thanks to the availability of the $1\,\rm deg \times 1\,\rm deg $ Megacam mosaic imager placed at the prime focus of the CFHT. Four one square degree fields are monitored every 3 to 4 nights during dark nights (i.e. for roughly 20 days centered on the new moon), in four filters $g'$, $r'$, $i'$ and $z'$. The discovery of new supernova candidates and the photometric followup of previous ones are obtained simultaneously with the same exposures. This presents the great advantage over previous generation experiments  of yielding a homogeneous set of supernova candidates all observed with a single instrument and with very good light curve coverage: several points in each filter are usually obtained before the light curve maximum. In addition, the availability of several filters ensures the coverage of the light curve in rest-frame $B$ band up to redshifts of $\sim 1$. This allows a better comparison of nearby and distant supernovae, improving the systematics on the cosmological parameters derived from the measurements. 

%\begin{table}[h]
%\begin{center}
%\begin{tabular}{|c|c|c|}
%\hline
%Field & RA (2000) & Dec (2000)\\
%\hline
%D1 & 02:26:00.00 & -04:30:00.0 \\
%D2 & 10:00:28.60 & +02:12:21.0 \\
%D3 & 14:19:28.01 & +52:40:41.0 \\
%D4 & 22:15:31.67 & -17:44:05.7\\
%\hline
%\end{tabular} \caption{Deep fields of the CFHTLS, used for the SNLS.}
%\label{tab:fields}
%\end{center} \end{table}

With this setup and the rolling-search mode, the SNLS expects to measure the matter density $\Omega_M$ and the dark energy density $\Omega_\Lambda$  of the Universe to better than 10\% and the equation-of-state parameter $w$ to $\sim 0.1$.

\section{Real time operations}

Starting from real-time preprocessed data (Magnier \& Cuillandre 2004), two independent real-time analysis pipelines (run by the French and Canadian teams) analyze the data as it arrives from the Mauna Kea. Both pipelines operate on the basis of image subtraction, after matching the point spread function of a given exposure to that of the image used as reference. This is done with the Alard (Alard \& Lupton 1998; Alard 2000) algorithm for the French team, and with a non-parametric approach (Pritchet 2005) for the Canadian team. After suppression of various artefacts due to saturated stars, plane tracks and cosmic rays for instance, a catalog of about ten new detections per CCD is obtained after every night of observation. A subsequent step of visual inspection of all the candidates leaves about five remaining candidates per mosaic and per night of observation, among which are selected those sent to spectroscopy (see the paper of S. Basa in the same proceedings for details on the spectroscopy program and results). 

A significant improvement over previous experiments or over the first months of the survey is obtained with the fit of the multi-color pre-maximum photometric data for each of the new candidates (Sullivan et al. 2005). This fit allows a good segregation between the various types of supernovae, thus enhancing the ratio of type Ia supernovae in the sample of candidates selected for spectroscopy. In addition, the fit outputs an estimation of the date of maximum and an approximate redshift, both used to optimize the quality of the spectrum by choosing the appropriate night (near maximum) and  the appropriate exposure time (increasing with increasing redshift). 

\begin{figure}[h]
   \centering
      \includegraphics[width=\textwidth]{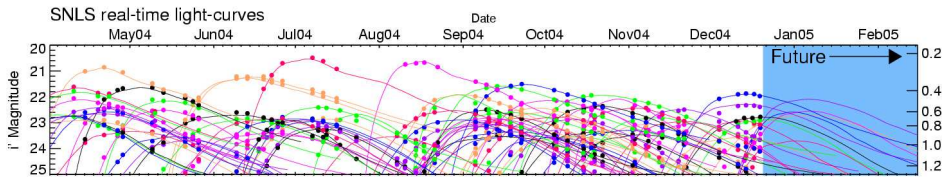}
      \caption{Light curves (brightness vs. time) in the $i'$ filter from SNLS data. It appears clearly that the maximum brightness, on which cosmological analysis depends, can be measured to very high accuracy from these data.}
       \label{fig:lc}
   \end{figure}

As of January 2005, $\sim 650$ supernova candidates have been detected, $\sim 230$ of which having spectral data and 150 being type Ia SN (cf. figure~\ref{fig:lc}).
%(cf. illustration of real-time light curves in figure~\ref{fig:lc}).

The main drawback of the current real-time analysis is the introduction of a step of visual inspection within the selection pipeline. Subjective selection biases can be introduced either in the selection of the valid candidates amongst the background of fake detections, or in the selection of the candidates to be sent to spectroscopy. An independent analysis is therefore performed afterwards, called offline selection and analysis, that aims  at discovering additional type Ia supernovae while measuring the selection biases of the real-time pipeline. 

\section{Offline selection and analysis}

This analysis is also based on image subtraction but unlike the real-time pipeline it builds full light curves (covering two years as of June 2005) of all the objects detected on any of the subtraction images. A search for SN candidates is then performed on these light curves, while their photometry is obtained with the SALT code (Guy et al. 2005). The offline search  includes three major requirements: 1) existence of a unique and significant fluctuation on the light curve. This excludes most photometric artefacts. 2) candidate not centered on a star, where the galaxy/star segregation of the hosts is based upon the width of the host on the reference image, up to a magnitude of $m_{i'} = 22.5$ where galaxies not only dominate largely but also can no longer be distinguished from stars. This excludes variable stars and some active galactic nuclei or quasars. 3) reasonable fit of the light curve by a generic shape of supernova (no constraint on absolute flux). 

The offline search  is expected to be sensitive to all supernovae above a given signal-to-noise and not be biased for instance against supernovae centered on its host galaxy (which can be confused with active galactic nuclei and have therefore sometimes been rejected in the real-time processing). The major improvement with respect to the real-time analysis is expected to occur for redshifts from 0.6 to $\sim$ 1.2. Therefore, the filter best adapted to this analysis is $i'$. The selection for spectroscopy end up in a cut on the magnitude of the candidate around $m_{i'} \sim 24.2$. An offline analysis going $0.5$ magnitude deeper than the real-time pipeline should lead to an additional set of about 40 supernovae per year. 

\begin{figure}[h] \centering
 \mbox{ \subfigure{\includegraphics[width=.45\textwidth] {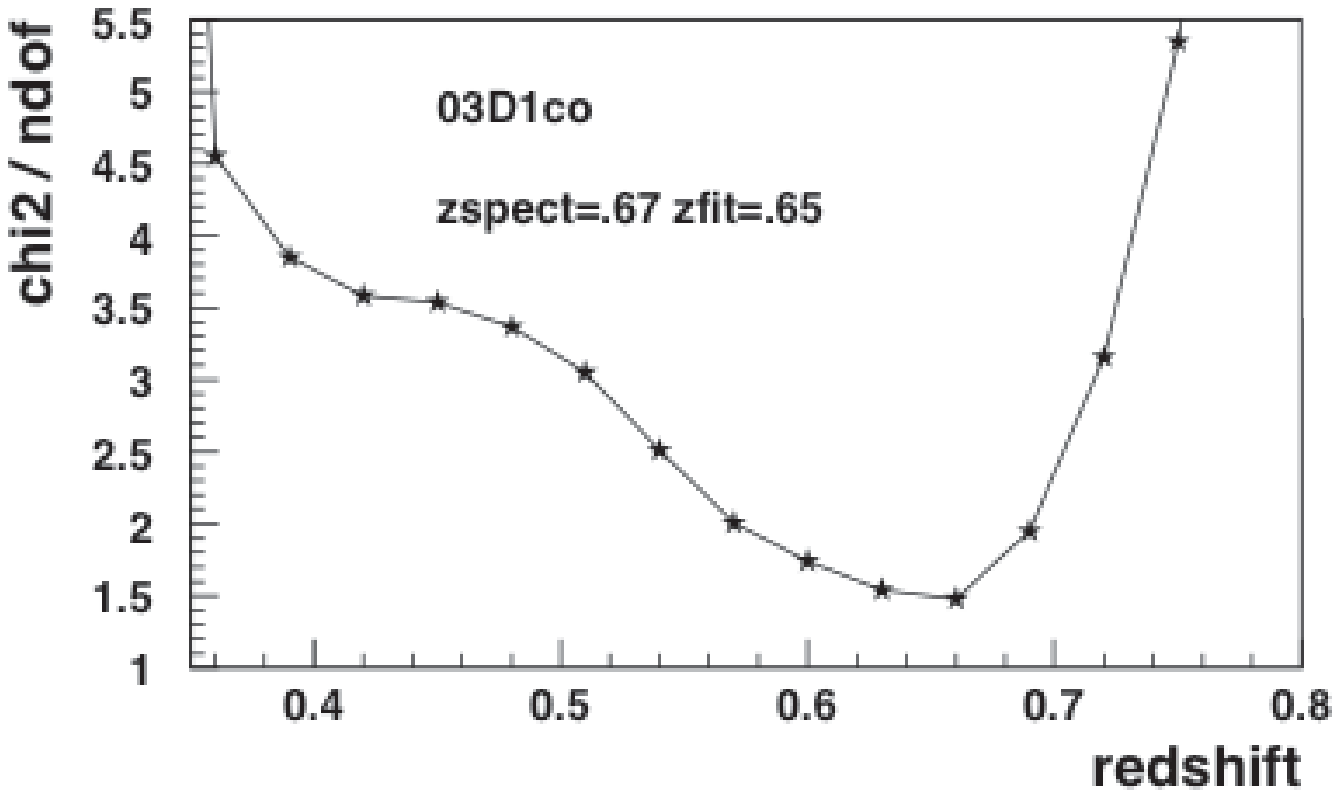}}
  \quad \subfigure{\includegraphics[width=.45\textwidth] {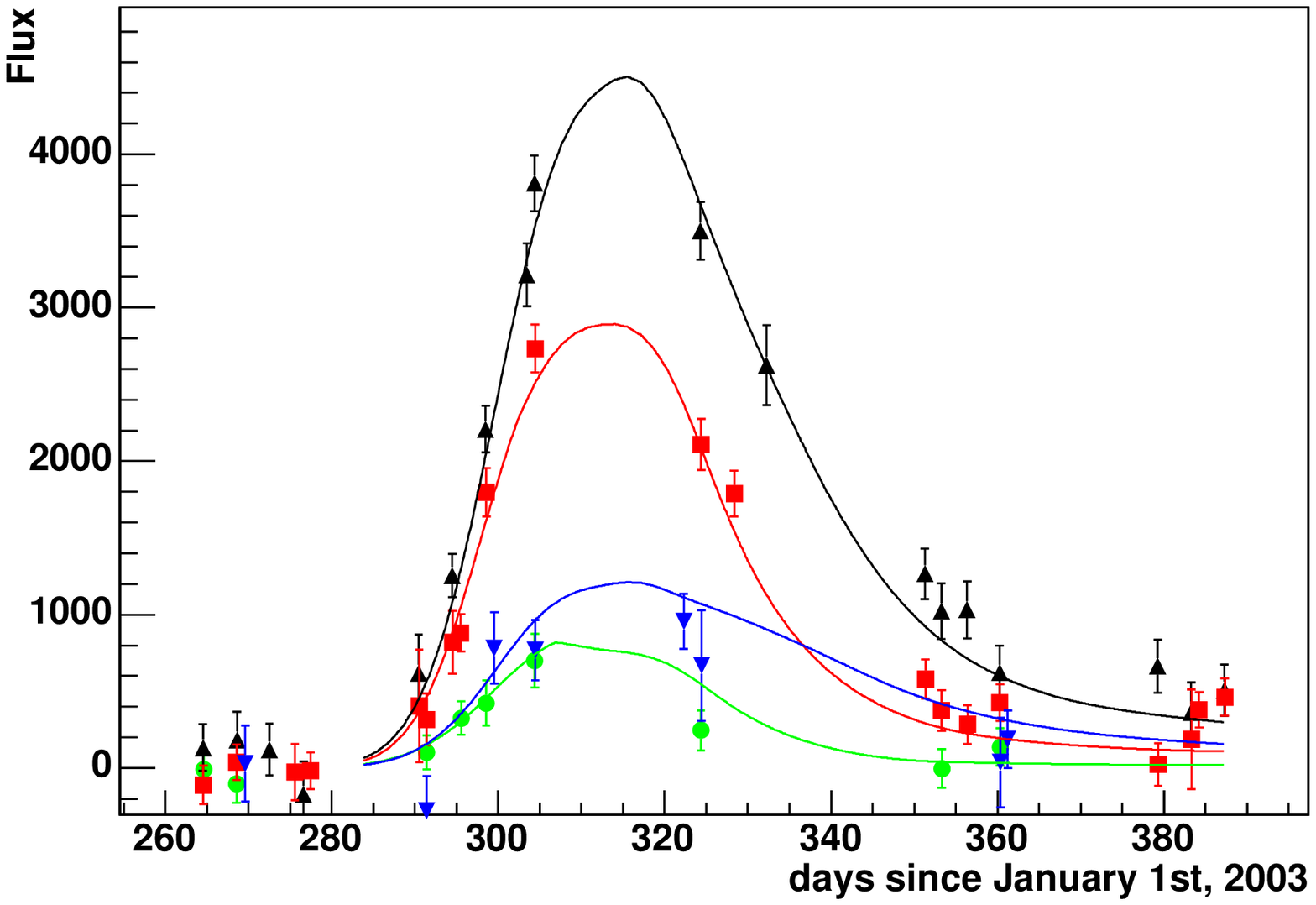}} }
  \caption{Left: $\chi^2$ map for the redshift determination of SN 03D1co. The photometric fit yields an estimate of the redshift compatible with spectroscopy. Right: 4-filter (from bottom to top: $g'$, $z'$, $r'$, $i'$) light curves at a redshift of 0.67.} \label{fig:z}
\end{figure}

For a cosmological use of the offline candidates, their redshift is required. In some cases, it will be possible to use later allocated spectroscopic time to measure the redshift of the galactic host of the supernova, once the supernova has faded away. In addition, a study is currently under investigation to determine the redshift from the 4-band photometric light curves alone, based on the fit of a SN Ia light curve template (SALT model).  Scanning the range of redshifts from 0 to 1.2, a $\chi^2$ map is built for each candidate, as illustrated in figure~\ref{fig:z}. For the first 75 real-time candidates, this determination of the redshift has a statistical uncertainty of  0.05 only for secure type Ia supernovae and of 0.10 for possible type Ia candidates.

\section{Conclusions}
The SNLS is progressing fast. The sample of confirmed type Ia supernovae already exceeds in number and in quality all previously published work on this topic. With three more years to go, the SNLS team expects to improve significantly the constraints on the cosmological parameters and on the equation-of-state.

%\subsection{ Literature citations}

% The following examples illustrate the required style in the main text:
% (Copernicus 1986); (Copernicus \& Galilei 1988); (Hubble et al. 1985;
% Newton et al. 1987; Ptolemaus \& Copernicus 1989a, 1989b, 1992);  Recently
% Galilei et al. (1992, 1993) showed that...

%%-----------------------------
%%      your bibliography
%%-----------------------------
%In preparing the reference list please adhere to the following format.
% Attention should be paid to the order of the items in each reference
% and to the punctuation used. Please see Sect. 4 in the User's Guide
% that comes with the new macro package.

%Bohr, N., Einstein, A., & Fermi, E. 1992, MNRAS, 301, 257 (BEF)
% Curie, M., & Curie, P. 1991, A&A, 248, 612
% de Gaulle, C. 1996, Solar Phys. (Oxford: Oxford Univ. Press)
% Heisenberg, W., & West, C. N. 1993, Australian J. Phys., 537, 36  (Paper III)
% Laurel, S., & Hardy, O. 1994, Active Galactic Nuclei, in The Evolution
% and Distribution of Galaxies, ed. W. Churchill, F. D. Roosevelt, & J.
% Stalin (New York: Wiley), 210


\begin{thebibliography}{}
\bibitem{}Riess, A. et al., 1998, AJ, 116, 1009
\bibitem{}Perlmutter, S. et al., 1999, ApJ, 517, 565 
\bibitem{}Magnier, E. A. \& Cuillandre, J.-C., 2004, PASP, 116, 449 
\bibitem{}Alard, C. \& Lupton, R. H. 1998, ApJ, 503, 325 
\bibitem{}Alard, C. 2000, A\&AS, 144, 363
\bibitem{}Pritchet 2005, in preparation
\bibitem{}Sullivan et al. 2005, submitted to AJ
\bibitem{}Guy et al. 2005, A\&A, in press
\end{thebibliography}
\end{document}